\documentclass[11pt,letterpaper]{article}
\usepackage[margin=0.7in]{geometry}
\usepackage[T1]{fontenc}
\usepackage{amsmath,amssymb}
\usepackage{graphicx}
\usepackage{booktabs}
\usepackage{array}
\usepackage[dvipsnames]{xcolor}
\usepackage[round]{natbib}
\usepackage[hidelinks]{hyperref}
\usepackage{titlesec}
\usepackage{enumitem}
\usepackage{wrapfig}
\usepackage{caption}
\usepackage{multicol}
\usepackage[most]{tcolorbox}
\usepackage{aas_macros}   
\titlespacing*{\section}{0pt}{6pt}{2pt}
\titlespacing*{\subsection}{0pt}{5pt}{2pt}
\titleformat{\section}{\normalfont\large\bfseries}{\thesection.}{0.5em}{}
\titleformat{\subsection}{\normalfont\normalsize\bfseries}{\thesubsection.}{0.5em}{}
\newcommand{\afe}{[$\alpha$/Fe]}
\newcommand{\feh}{[Fe/H]}
\newcommand{\jwst}{\textit{JWST}}
\newcommand{\hst}{\textit{HST}}
\newcommand{\micron}{\mu\text{m}}
\newcommand{\sqam}{\,arcmin$^{2}$}
\newif\ifcoverpage
\coverpagetrue

\title{\ifcoverpage\vspace*{1.8in}\else\vspace{-28pt}\fi\Large\bfseries A Multi-Cycle Treasury Survey of Andromeda:\\ Comprehensive, All-Encompassing Spectroscopy, Imaging, and Mid-Infrared Mapping of the Nearest Giant Spiral\vspace{-4pt}}
\author{\parbox{0.92\textwidth}{\centering\normalsize
David M.\ Nataf (University of Iowa)\thanks{Corresponding authors: David Nataf,
\href{mailto:david-nataf@uiowa.edu}{david-nataf@uiowa.edu},  and David Nidever, \href{mailto:david.nidever@montana.edu}{david.nidever@montana.edu}.
\textit{Generative AI:} design code and parts of the text were drafted with Claude; the science case, numbers, and
conclusions are the authors'.}
and David L.\ Nidever (Montana State University)$^{*}$,\\
Gagandeep Anand, Magda Arnaboldi, Joss Bland-Hawthorn, Scott Call, Santi Cassisi, Marcio Catelan, Julianne Dalcanton, Ivanna Escala, Mario Gennaro, Benjamin Gibson, Karoline Gilbert, Michael Hayden,
Danny Horta, Caroline Huang, Rodrigo Ibata, Christian Johnson, Evan Kirby, Chiaki Kobayashi, Andrea Kunder, Geraint Lewis, Anna Liptrott, Massimo Marengo, Melissa Ness, Rachel Pauline, Henry Prager, Zdenek Prudil, Henrique Reggiani, Michael Rich, Ricardo Schiavon, Sanjib Sharma,
Brian Thorsbro, Yuan-Sen Ting, Dan Weisz, Rosemary Wyse, Gail Zasowski, and Catherine Zucker}\vspace{-6pt}}

\begin{document}

\maketitle
\thispagestyle{empty}
\ifcoverpage
  \clearpage            
  \setcounter{page}{1}  
  \pagestyle{plain}
\else
  \pagestyle{empty}
\fi

\begin{tcolorbox}[enhanced, colback=Lavender!25, colframe=Lavender!70!black, boxrule=0.4pt, arc=1.5pt,
  left=6pt, right=6pt, top=4pt, bottom=4pt, boxsep=0pt, borderline west={2.5pt}{0pt}{Lavender!70!black}]
\small
{\normalsize\bfseries Abstract}\quad
We propose a Multi-Cycle Treasury program of $\sim$800 hours that would survey Andromeda the way the Milky Way has been surveyed by
APOGEE, Gaia, and WISE.
\begin{itemize}[nosep, leftmargin=1.4em, itemsep=1pt]
  \item \textbf{NIRSpec:} spectra of $\sim$4,500 red giants and supergiants at S/N $\approx$ 80, sufficient for 16 elements at
        $\lesssim$0.05 dex precision and radial velocities to a few km\,s$^{-1}$.
  \item \textbf{NIRCam:} $\sim$1,600\sqam\ in up to ten filters, reaching $\sim$20 million stars, every field imaged twice a year
        apart: two-epoch infrared photometry of the $\sim$500 Cepheids and $\sim$10$^{4}$ Miras in the footprint.
  \item \textbf{MIRI:} $\sim$450\sqam\ in five filters at the depth of PHANGS.
\end{itemize}
The observations are predominantly organized as twelve ``bricks'', each of three adjacent fields including two NIRSpec/MOS pointings, in which one instrument is prime and a second observes
in parallel, so that roughly 40\% of the exposure time comes at no charge. The bricks sample the nucleus, bulge, major axis to 29 kpc,
minor axis, and inter-arm disk. In addition, a 100-hour grid of short NIRCam pointings with MIRI in parallel covers the \textit{HST}-PHAT
footprint. The result would be the first chemodynamical, star-formation, and dust map of an $L^{\star}$ galaxy other than our own
built from individual stars, at a scale no single-cycle program can reach.
\end{tcolorbox}

\section{Scientific justification}

\begin{wrapfigure}{r}{0.6\textwidth}
\vspace{-5pt}
\centering
\includegraphics[width=0.59\textwidth]{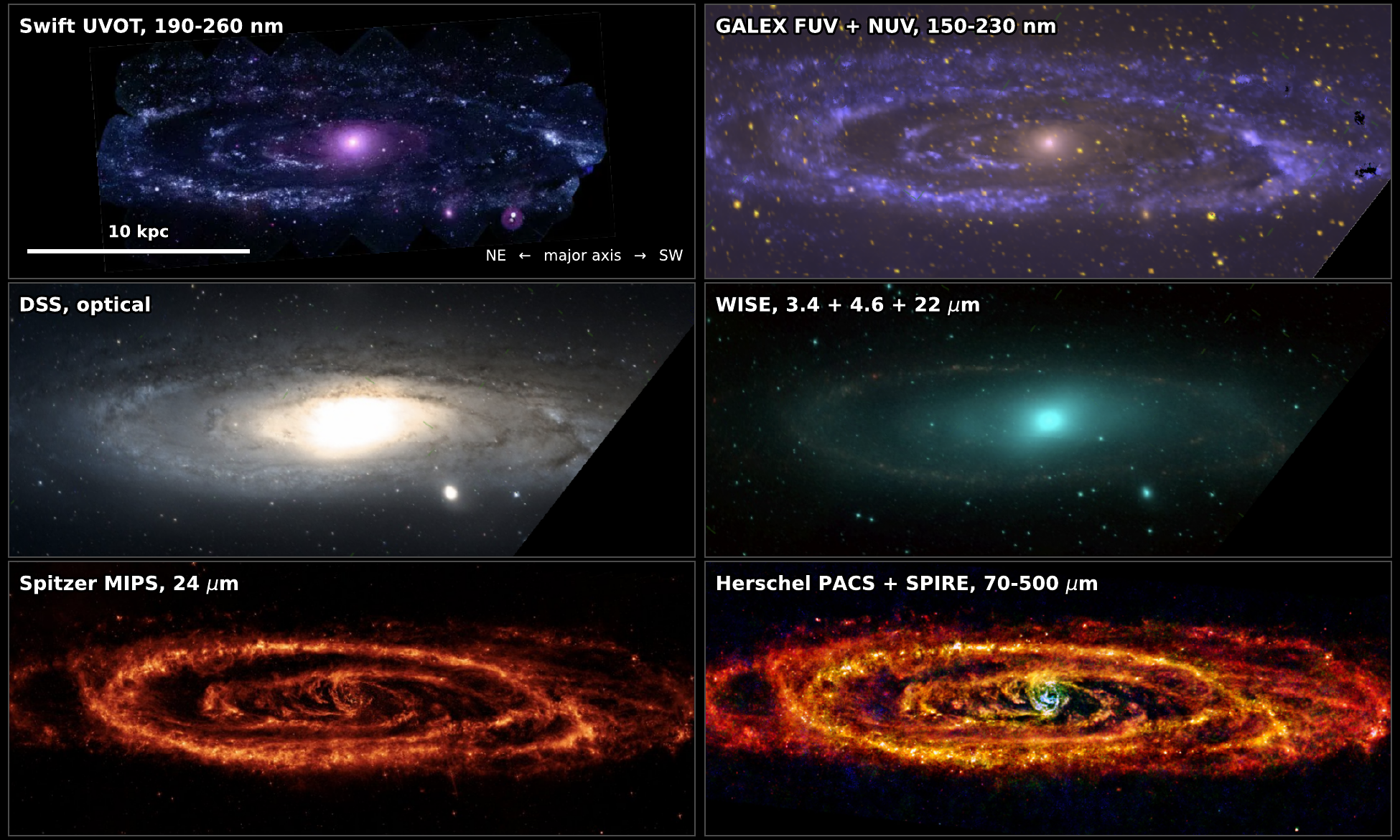}
\caption{Andromeda from the ultraviolet to the far-infrared on one $2.3^\circ \times 0.9^\circ$ frame, major axis horizontal,
north-east to the left: Swift/UVOT and GALEX (young stars), DSS (the old disk and bulge that NIRSpec and NIRCam sample), WISE
3.4--22\,$\mu$m, Spitzer/MIPS 24\,$\mu$m (warm dust and PAHs, the MIRI bands), Herschel PACS+SPIRE (cold dust). Credits:
NASA/Swift/S.~Immler; NASA/JPL-Caltech; STScI/DSS; ESA/Herschel/J.~Fritz.}
\label{fig:multiwavelength}
\end{wrapfigure}

\textbf{Andromeda is an $L^{\star}$ galaxy like the Milky Way, but quite distinct from the Milky Way.} M31 (Fig.~\ref{fig:multiwavelength}) is often called the Milky
Way's twin, and it is a disk galaxy of the same Hubble type and luminosity class. There are key differences. It carries about twice
the stellar mass \citep[$\sim 10^{11}\,M_\odot$;][]{2015IAUS..311...82S,2015ApJ...806...96L} yet forms stars at a fraction of the
Milky Way's rate \citep{2015ApJ...805..183L,2015ApJ...806...96L}. It had a major merger $\sim$2 Gyr ago, whose debris is the Giant
Stellar Stream and whose remnant may be M32 \citep{2018NatAs...2..737D,2018MNRAS.475.2754H}. It hosts $\sim$450 globular clusters
against our $\sim$180 \citep{2016ApJ...824...42C,1996AJ....112.1487H}. Its central black hole is 20 to 30$\times$ more massive than
Sgr A$^{*}$ \citep{2005ApJ...631..280B,2022A&A...657L..12G}, and is inside a composite bulge whose classical component the Milky Way lacks
\citep{2018MNRAS.481.3210B}. Finally, its center emits gamma rays at \textit{eight times the rate} of a Milky Way scaling
\citep{2019ApJ...871L...8F}. Everything the Milky Way has taught us about how disk galaxies assemble was learned from one object with
one history. Andromeda is the one other galaxy of this class whose assembly can be read the same way, star by star, and only \jwst\ can reveal it.

\textbf{What the stars record.} Core-collapse supernovae enrich the interstellar medium with $\alpha$ elements within a few $10^{7}$ yr of a star
formation event, whereas Type Ia supernovae supply most of the iron over $10^{8} - 10^{9}$ yr timescales
\citep{1979ApJ...229.1046T,1986A&A...154..279M}. A population that formed quickly is $\alpha$-enhanced at every metallicity, one
that formed slowly turns over toward solar ratios once Type Ia enrichment catches up. This clock, read from \afe\,
yields the prediction that bulges and ellipticals formed in early, rapid bursts \citep{1990ApJ...365..539M}, which the first high-resolution
spectra of bulge giants bore out \citep{1994ApJS...91..749M}, and it is why the two sequences of Milky Way disk stars in the
\afe--\feh\ plane, the higher-\afe\ one older and thicker \citep{2015ApJ...808..132H,2017AJ....154...94M}, are read as separate sequences of star formation and chemical enrichment.

Nearly all of this rests on one galaxy, the Milky Way. Spectroscopic surveys and Gaia have given abundances, ages, and orbits for millions of Milky
Way stars \citep{2016ARA&A..54..529B}: an old, $\alpha$-rich thick disk and a younger, $\alpha$-poor thin disk,  a bar-dominated bulge
whose abundances follow the thick disk \citep{2018ARA&A..56..223B} and whose age is still debated
\citep{2008ApJ...684.1110C,2017A&A...605A..89B,2018ApJ...863...16R,2024MNRAS.530.2972S}, and a halo assembled from identifiable
accretion events. But one galaxy is but one galaxy, and this limits our ability to infer general statements about how galaxies form and evolve. The $\alpha$-bimodality is a clear example:
\citet{2018MNRAS.477.5072M} found it in only 6 of 133 Milky Way-mass galaxies in the EAGLE simulations and \citet{2026arXiv260102520B} in 4 of 16 in the
FIRE simulations, a prediction that it is not a generic feature, but plausibly an outcome of histories where the star-formation rate had a particularly rapid decline. In contrast, other theorists have argued it may be the outcome of a gas-rich merger \citep{2020MNRAS.491.5435B}, whereas
\citet{2009MNRAS.396..203S} and \citet{2019MNRAS.484.3476C} obtain it from (more generic) radial migration or clumpy star formation, in which case
it should be the norm. Bulges carry the same ambiguity: models of a classical bulge formed in a $\sim$0.5 Gyr burst predict $\alpha$
enhancement to the highest metallicities \citep{2007A&A...467..123B,2015MNRAS.447.1096M}, but the Milky Way
bulge does not look that way, it appears to contain a trace classical bulge at most. The stars of ellipticals, where these prediction matters most, cannot be resolved. The bulge of Andromeda is our best hope to understand classical bulges, and \jwst~is the only observatory that can effectively study it. Among other issues, we are curious to know if the classical bulge stars of Andromeda show much higher relative abundance ratios of [$\alpha$/Fe], [Na/Fe], and [N/Fe] and lower values of [C/Fe] and [O/Fe] than the stars in the Milky Way Bulge, as expected by some models and observations of ellipsoidal formation \citep{2017ApJ...841...68V,2018ApJ...857...16G,2019MNRAS.483.3420P}.

\begin{wrapfigure}{r}{0.69\textwidth}
\vspace{-7pt}
\centering
\includegraphics[width=0.68\textwidth]{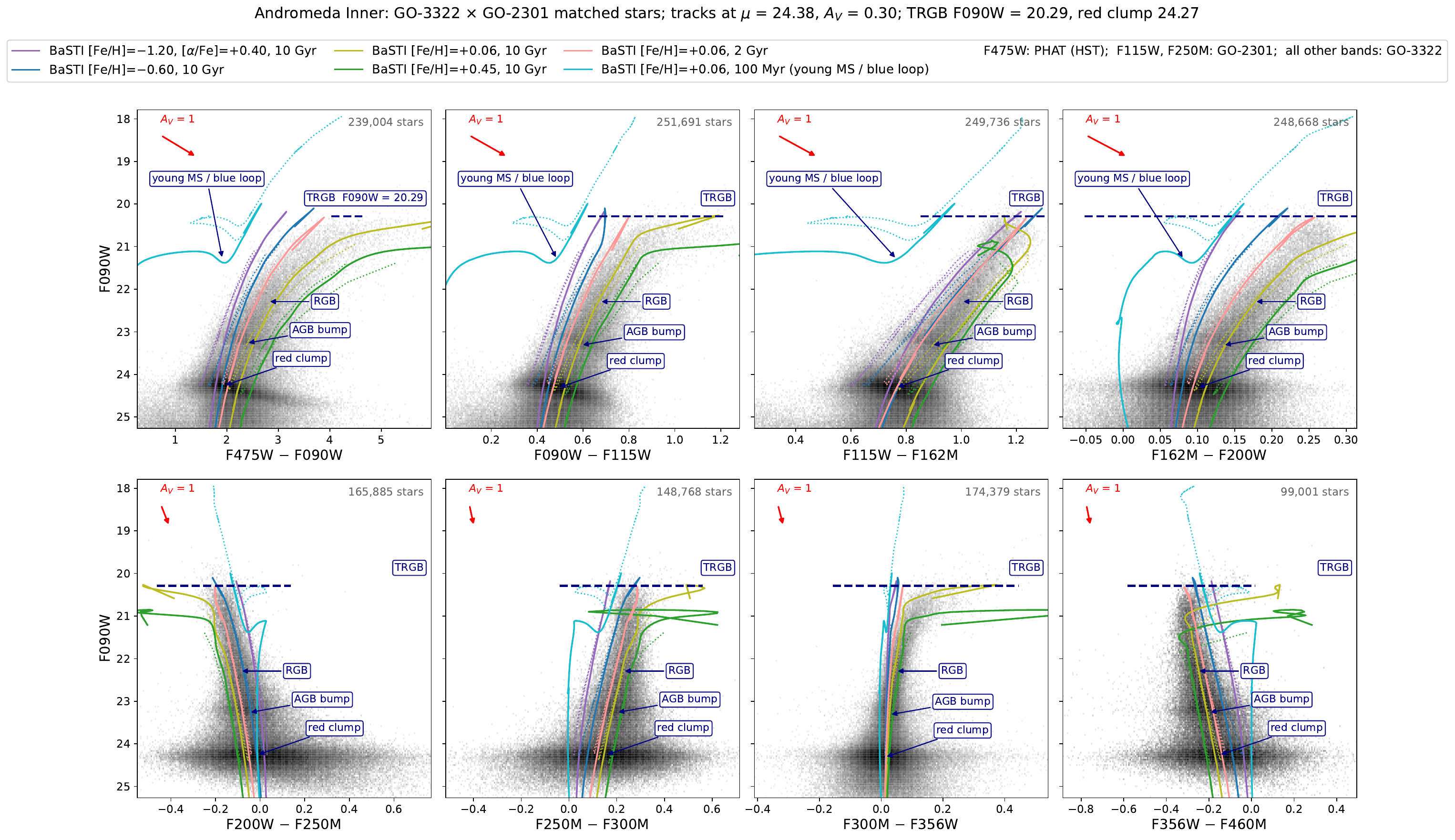}
\caption{Color-magnitude diagrams of an Andromeda disk field from archival GO-2301+3322 
photometry in eight NIRCam filters (as well as archival \hst $F475W$), with BaSTI isochrones at $\mu = 24.38$, $A_V = 0.30$. The tip of the red giant branch, AGB bump, and red clump are resolved in every color. The range in wavelengths can distinguish colder temperatures from higher extinctions, and the medium-band filters can distinguish greater \feh from greater \afe, see Figure~\ref{fig:M31}}
\label{fig:cmd}
\end{wrapfigure}
Andromeda's inclination shows us its bulge and nucleus through a fraction of the extinction that hides our own. Integrated light
finds super-solar \feh\ and \afe\ in the bulge \citep{2018A&A...618A.156S,2021A&A...647A.131G,2025arXiv251115415L} and planetary
nebulae trace the disk gradient \citep{2019A&A...631A..56B,2023ApJ...956L..14K}, but star-by-star abundances were until recently out
of reach: Keck spectroscopy of M31 halo giants reaches \afe\ only by coadding many spectra \citep{2023ApJ...951...12W}. \jwst/NIRSpec has
changed this. In one MSA field at 18 kpc on the major axis, GO-2609 obtained spectra of $\sim$100 red giants at a mean S/N of 85,
precisions of 0.05 dex in \feh\ and \afe, and of 8 km\,s$^{-1}$ in radial velocities, and found a single $\alpha$-rich sequence
with no analogue to the Milky Way's low-$\alpha$ disk \citep[][Fig.~\ref{fig:spectrum}]{2024IAUS..377..115N}. The result fits the
thick, hot disk that PHAT and SPLASH found \citep{2015ApJ...803...24D,2023AJ....166...80D}, but it rests on one sightline, and the
other M31 NIRSpec programs (GO-2301 and GO-4735) will raise the total only to $\sim$600 stars, all
along the major axis and none within the Bulge, the minor axis, or in the outer disk. We do not know whether the single high-$\alpha$ sequence holds throughout Andromeda; we do not know whether an age--velocity relation like the Milky Way's
\citep{2019MNRAS.489..176M} exists in Andromeda; We do not know the Andromeda disk's stellar abundance gradients in \feh\ and \afe\. Answering these questions requires analogous measurement to the ones that produced the Milky Way picture, applied across Andromeda. At this time, these abundance measurements can only be achieved with \jwst's NIRCam and NIRSpec, and the kinematic measurements can only be achieved with NIRSpec.

\section{Scientific scope of the survey}

\textbf{Chemical evolution.} With $\sim$4,500 stars at S/N $\approx$ 80 in G140H and G235M we expect to measure C, N, O, Mg, Si,
Ca, Ti, Al, Na, K, Fe, Mn, Cr, Co, Ni, and Sr (Table~\ref{tab:lines}) to $\lesssim$0.05 dex per star \citep{2024IAUS..377..115N},
exceeding the precision necessary to distinguish two sequences analogous to the Milky Way's, in a sample large enough to ask the question separately in the bulge,
inner and outer disk, minor axis, and halo. 

\begin{wrapfigure}{r}{0.42\textwidth}
\vspace{-10pt}
\centering
\includegraphics[width=0.41\textwidth]{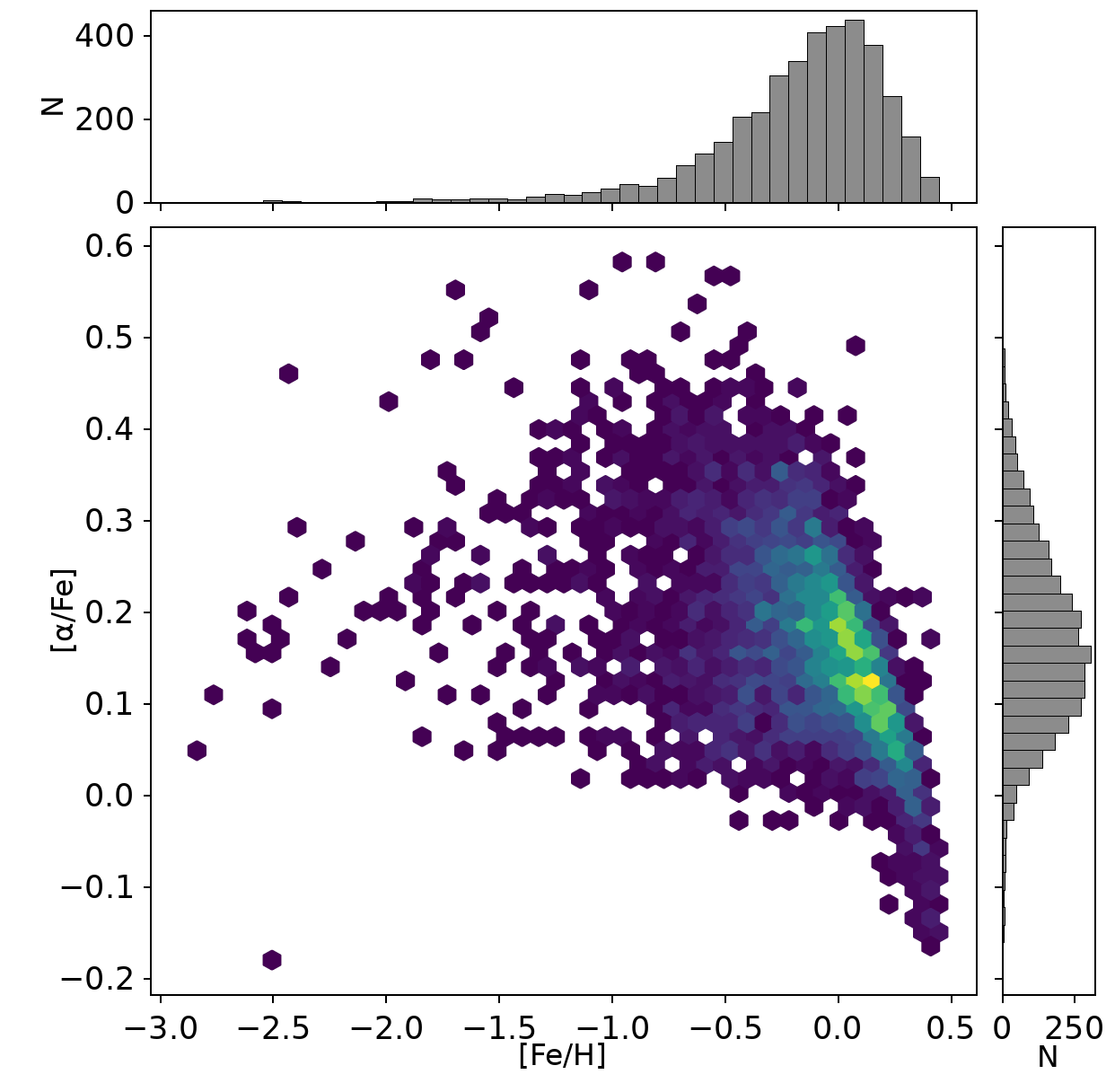}
\caption{Proof of method: \afe\ against \feh\ for upper-RGB stars in the outer disk of M31 from our Bayesian isochrone fits to
archival NIRCam photometry (GO-2609, GO-3322), with marginal distributions. Fewer bands, shallower exposures, and fewer dithers than
proposed here already resolve a metal-rich concentration near solar metallicity and a metal-poor, $\alpha$-richer tail.}
\label{fig:M31}
\end{wrapfigure}

\textbf{Dynamics.} Radial velocities to a few km\,s$^{-1}$ for the same stars, with ages from [C/N], give the first age--velocity
dispersion relation outside the Milky Way, from the nucleus to 29 kpc.

\textbf{Star-formation history.} The NIRCam mosaics reach the red clump in ten filters from 0.9 to 4.6\,$\micron$ over 1,600\sqam, extending the PHAT star-formation histories \citep{2015ApJ...805..183L,2017ApJ...846..145W}
into the infrared and beyond the PHAT footprint. The spectra calibrate photometric \feh\ and \afe\ for the $\sim$20 million imaged
stars in and of their own, with the benefit of calibration overlap on the $\sim$100,000 red giants imaged in $F390W$ by \hst\ program GO-18406, and the NIRSpec observations themselves. Our Bayesian isochrone
fits to the archival NIRCam photometry already resolve the \afe--\feh\ distribution of the outer disk (Fig.~\ref{fig:M31}). The
greater number of filters and the deeper exposures proposed here sharpen the sensitivity to extinction and blending, and so to \feh,
\afe, and age.

\textbf{Dust.} MIRI imaging in F560W, F770W, F1000W, F1130W, and F2100W at PHANGS-\jwst\ depth \citep{2023ApJ...944L..17L} maps
the PAH and warm-dust emission that Herschel and Spitzer see only at coarser resolution (Fig.~\ref{fig:multiwavelength}) at 1--3 pc
resolution over 450\sqam; with the NIRCam colors of the bright giants it gives the corresponding dust extinction curve and normalization.

\textbf{Variable stars.} Every field is imaged twice, a year apart. The footprint holds $\sim$500 Cepheids, 371 of them with PHAT
photometry \citep{2015ApJ...799..144K} out of $\sim$2,700 known across the disk \citep{2018AJ....156..130K}, and of order $10^{4}$
Miras. Each receives two-epoch NIRCam photometry in up to ten filters, and the Miras and Cepheids in the brick fields receive NIRSpec spectra: the
spectroscopically anchored O-rich/C-rich classification \citep{2013ApJ...774...83B,2024MNRAS.531..110K} that the Mira distance
ladder requires \citep{2020ApJ...889....5H}, and metallicities for individual Cepheids beyond the Magellanic Clouds. Our imaging strategy includes the filters $F182M$, $F300M$, and $F460M$, which enable determinations of carbon monoxide and water abundance, and also distinguishing between oxygen-rich and carbon-rich atmospheres \citep{2016MNRAS.459.1170S,2026arXiv260930551H}. Multiple epochs are beneficial due to the phase dependence of some of these features. 

\textbf{The halo.} The minor-axis and outermost bricks reach the inner halo, where \citet{2026arXiv261024915H} have photometric \feh\ and \afe\ for 76,400 Subaru-HSC giants calibrated on 19 spectra; ours would calibrate the whole map to sightlines with much higher stellar densities.

\section{Why a Multi-Cycle Treasury program}

Three things put this survey beyond single-cycle programs. The first is scale: the NIRSpec programs executed and approved so far will total
$\approx 600$ stars after five cycles, and a 4,500-star sample at S/N 80 requires 370 h of NIRSpec time before any imaging. The second is geometry: the MSA configurations, the NIRCam pre-imaging that selects their targets, and the MIRI parallels line up only at the roll angles of
M31's winter window, and spectroscopy must follow its pre-imaging by a minimum of 60 days, so each field's imaging, spectroscopy, and MIRI epochs
span three winters however the program is sized. And the third reason is uniformity: the archival programs differ in selection, depth, and gratings;
comparing bulge, disk, and halo at 0.05 dex needs one selection, one exposure plan, and one reduction for every field.

\begin{figure}[tb]
\centering
\includegraphics[width=0.74\textwidth]{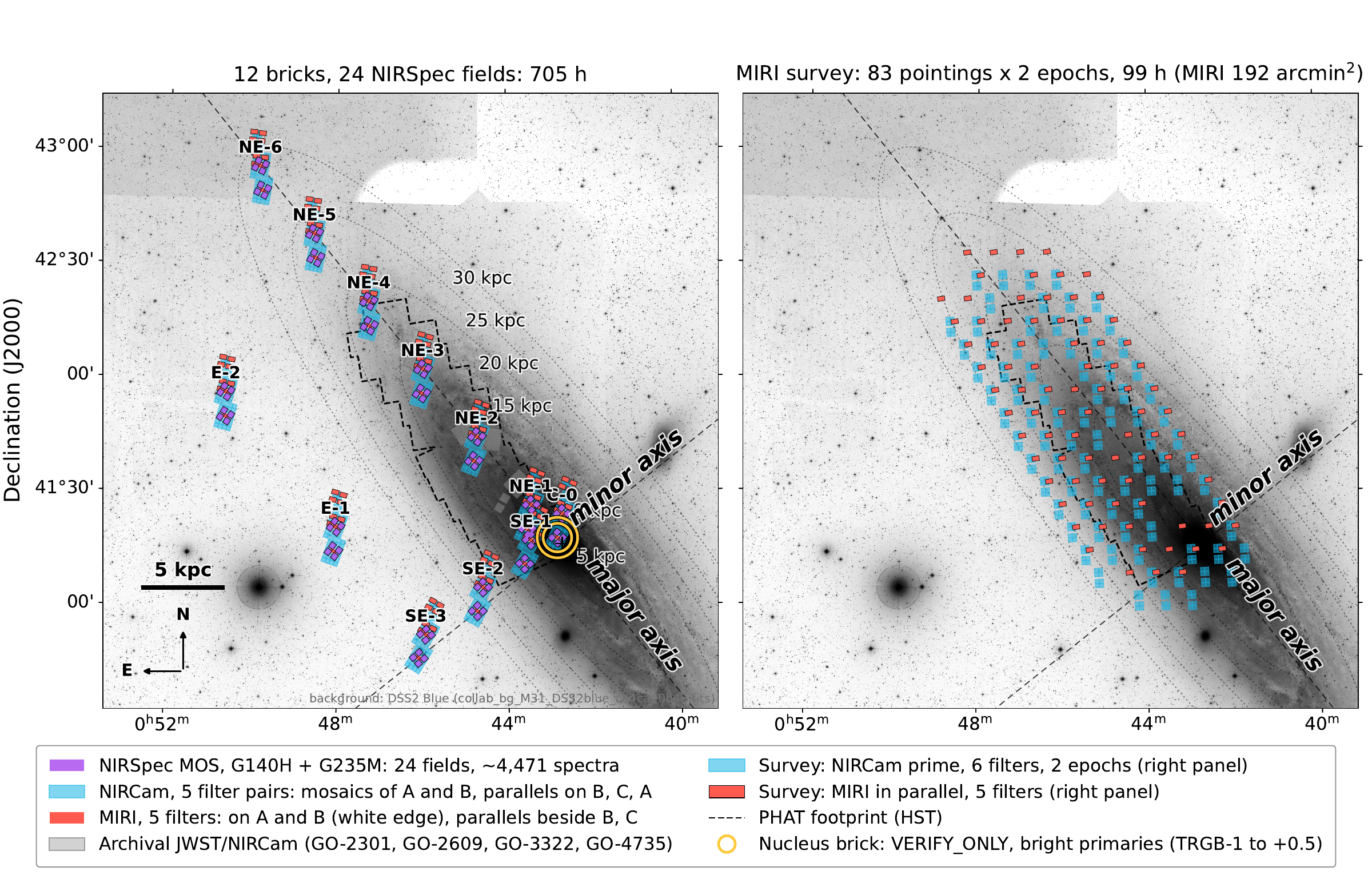}
\caption{The survey layout. Left: twelve bricks (NIRSpec in purple, NIRCam in blue, MIRI in red) at the nucleus, along the
north-east major axis to 29 kpc (NE-1 to NE-6), along the far-side minor axis (SE-1 to SE-3), and on a line parallel to the major
axis through SE-3 (E-1, E-2). Right: the MIRI-survey grid, 83 short
NIRCam pointings (blue) with MIRI in parallel (red), inside PHAT. Almost all MIRI pointings will land on a sightline with a prior or subsequent NIRCam pointing, to enable jointly mapping dust extinction and emission.}
\label{fig:layout}
\end{figure}

\section{Survey design}

The survey is built of two unit classes.  The first is a ``\textbf{brick}" of three fields A, B, and C in a row along the NIRSpec-to-NIRCam offset
(386$''$), observed over three winters: (1) NIRCam images A and B as $2\times2$ mosaics in five filter pairs (F090W/F250M,
F115W/F300M, F150W/F356W, F182M/F410M, F210M/F460M), covering 95\% of each NIRSpec MSA, with MIRI in parallel beside B and C; (2) NIRSpec
observes A while NIRCam, in parallel, images B again; (3) a year later NIRSpec observes B while NIRCam images C; (4) MIRI images A
and B at PHANGS depth, but in five filters rather than three. A standard brick costs 58 h charged and returns up to $\sim$500 stars with spectra, 67\sqam\ of NIRCam in ten filters (31\% imaged twice), and
22\sqam\ of MIRI. In all NIRSpec MOS pointings, Cepheids and Miras will be prioritized for targeting with 25$\times$ greater weight. 

\textbf{Layout.} Twelve bricks (Fig.~\ref{fig:layout}, left): two in the nucleus (0.5 kpc), six along the major axis at 3, 8, 13, 19, 24,
and 29 kpc, three along the far-side minor axis at 2.5, 6.5, and 11 kpc projected, and one off both axes. Ten fields lie outside
PHAT \citep{2023ApJS..268...48W}. All targets will be selected from the program's own NIRCam pre-imaging.

\textbf{NIRSpec.} G140H/F100LP ($R\sim2700$) and G235M/F170LP, with three MSA configurations dithered in the dispersion direction by 0, 80,
and 160 shutters to bridge the detector gap with redundancy, and sub-shutter offsets of $\pm 0.125$ shutters to recover full PSF sampling \citep{2024IAUS..377..115N}. Primaries
are drawn from the top magnitude of the giant branch (F160W from the tip to 1 mag below). Fillers extend 1 mag fainter, include all brighter M31 stars, as well as the OB stars of comparable magnitudes. At 15.2 h charged per field the ETC gives S/N $=78$ per pixel 1.4 mag below the tip
(F814W $=21.5$) and $>$100 at the tip; MPT places $\sim$250 stars per field in two or three of the three configurations.

\textbf{MIRI survey.} Eighty-three short NIRCam pointings (F090W/F250M, F115W/F356W, F182M/F460M, 258 s each) on a hexagonal grid
placed so that the MIRI parallels tile the PHAT footprint without overlap (Fig.~\ref{fig:layout}, right), walked twice with
different MIRI filters, so that every MIRI pointing ends with five filters and every NIRCam pointing with two epochs in six filters,
for 99 h. Nearly all MIRI pointings will land on sightlines with both a NIRCam pointing and an archival \hst pointing from the PHAT survey, which includes ultraviolet imaging in $F275W$ and $F336W$. This would be the most exquisite joint mapping of dust extinction and emission of a galaxy other than the Milky Way.

\begin{table}[tb]
\centering\footnotesize
\caption{The program in numbers; charged times are calibrated on the APT file of one brick (NE-2, 58.5 h).}
\label{tab:budget}
\begin{tabular}{@{}lrrl@{}}
\toprule
Component & Visits & Hours & What it returns \\
\midrule
NIRCam $2\times2$ mosaics of A, B; MIRI parallel & 24 & 312 & 830\sqam\ in 10 filters; 256\sqam\ of MIRI beside B, C \\
NIRSpec on A, B; NIRCam parallel & 24 & 368 & $\sim$4,500 stars in G140H+G235M; B imaged again, C imaged \\
MIRI on A, B at PHANGS depth & 24 & 24 & 110\sqam\ in 5 filters on the spectroscopic fields \\
MIRI survey: NIRCam prime, MIRI parallel & 166 & 99 & 747\sqam\ in 6 filters, 2 epochs; 192\sqam\ of MIRI \\
\midrule
Total, three cycles, winter windows only & 238 & 804 & NIRCam 1,560\sqam, MIRI 450\sqam, 24 MSA fields \\
\bottomrule
\end{tabular}
\end{table}

\section{Feasibility}

\begin{figure}[tb]
\begin{minipage}[t]{0.57\textwidth}\vspace{0pt}
\centering\scriptsize
\setlength{\tabcolsep}{3pt}\renewcommand{\arraystretch}{0.95}
\captionof{table}{Unblended lines per element in a tip-of-the-giant-branch spectrum at $R=2700$ for the low-wavelength grating and $R=1000$ for the long wavelength grating, with depth $\geq 5\sigma$ per pixel
at S/N 80 and purity $\geq 0.7$. The metal-poor stars have fewer usable lines, due to both lower metallicity and higher temperature at fixed brightness.}
\label{tab:lines}
\begin{tabular}{@{}lccccl@{}}
\toprule
 & \multicolumn{2}{c}{G140H/F100LP} & \multicolumn{2}{c}{G235M/F170LP} & \\
Element & [M/H] $=0$ & $-1$ & [M/H] $=0$ & $-1$ & Traces \\
\midrule
Fe & 127 & 91 & 48 & 22 & \feh, Type Ia enrichment \\
Si, Ti & 29, 11 & 29, 10 & 21, 3 & 8, 2 & $\alpha$: core-collapse yields \\
Mg, Ca & 8, 2 & 6, 2 & 3, 10 & 2, 5 & $\alpha$: core-collapse yields \\
CN bands & 10 & 10 & -- & -- & [C/N] ages, mixing \\
CO bands & 1 & 1 & 18 & 16 & C, O, $^{12}$C/$^{13}$C \\
C & 5 & 2 & 4 & 1 & C/O, AGB dredge-up \\
Ni, Mn & 11, 9 & 5, 8 & 5, -- & 1, -- & Type Ia yields vs.\ metallicity \\
Cr, Co & 4, 2 & 1, -- & 1, -- & -- & Type Ia, hypernova yields \\
Al, Na & 3, 3 & 3, 2 & 3, 1 & 1, 1 & odd-$Z$, cluster-like enrichment \\
K, S, Sr & 1, 1, 2 & 1, 1, 2 & -- & -- & odd-$Z$, $\alpha$, $s$-process (Sr) \\
\bottomrule
\end{tabular}
\end{minipage}\hfill
\begin{minipage}[t]{0.41\textwidth}\vspace{0pt}
\centering
\includegraphics[width=0.9\textwidth]{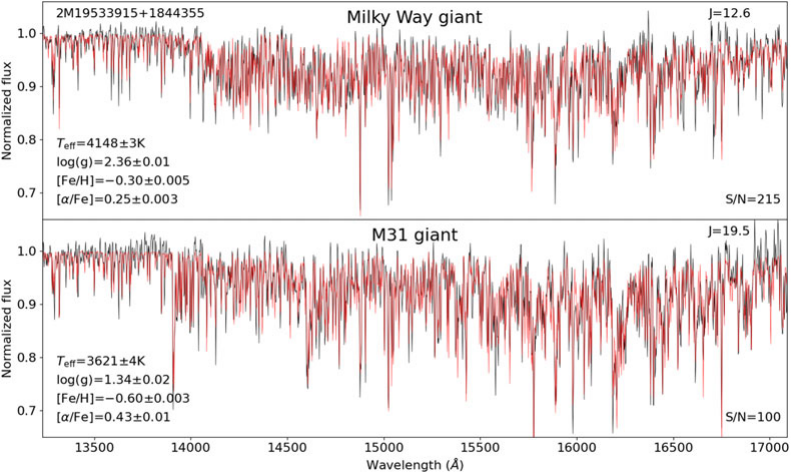}
\caption{NIRSpec G140H/F100LP spectra of an M31 giant at $J=19.5$ (bottom) and a Milky Way giant seven magnitudes
brighter (top), with best-fit synthetic spectra in red, from GO-2609; reproduced from \citet[][CC BY 4.0]{2024IAUS..377..115N}.}
\label{fig:spectrum}
\end{minipage}
\end{figure}

\textbf{Spectra.} GO-2609 showed that G140H/F100LP on the MSA yields S/N 70--100 spectra of M31 giants seven magnitudes fainter
than Milky Way calibrators (Fig.~\ref{fig:spectrum}), and Monte Carlo tests on \texttt{Synspec} grids give $\sim$0.05 dex for 15
elements at S/N $=70$ \citep{2024IAUS..377..115N}. Our MARCS-based synthetic spectra of M giants (3750--4000 K), convolved to
$R=2700$ and searched for isolated lines detectable at 5$\sigma$ per pixel at S/N 80, give the census in Table~\ref{tab:lines}. G235M
adds the CO band heads at 2.3 $\micron$, the only route to $^{12}$C/$^{13}$C, and the K-band Na, Ca, Mg, and Al lines, for 30\% of the NIRSpec time. These are conservative estimates as the resolution of NIRSpec's MOS has been shown to exceed design requirements. 

\textbf{Targets and crowding.} The MSA requires isolated stars. From PHAT v3 we count 1.7 million giants in the primary box; after
shutter-cleanliness cuts, the fields between 7 and 15 kpc keep $\approx 2000$ candidates each, while the nucleus and 3-kpc fields keep $\approx 500$ and use the bright-star regime with slitlet stepping in \texttt{VERIFY ONLY} mode. The predicted yields are 250 per field between 8 and 24 kpc, 170 in the nucleus and 3-kpc fields, and 100 in the
halo fields, $\sim$4,500 in all.  The NIRCam imaging from JWST-GO-12720, the first of the M31 bulge, is scheduled for August 2027, and will give us empirical requirements on how to deal with crowding in that regime, \textit{e.g.} how many additional dithers are required. 

\textbf{Scheduling.} The program executes in the winter visibility windows (V3PA 82$^\circ$--104$^\circ$), outside the
micrometeoroid avoidance zone, at 230--310 h per cycle (15--21\% of each two-month window). We have validated that a one-brick APT file passes the
Visit Planner and the NIRCam+MIRI data-rate limit.

\textbf{Observing time and legacy.} The program requires 804 h over three cycles (Table~\ref{tab:budget}). The products are a
catalog of $\sim$4,500 stars with abundances, velocities, and ages; ten-band photometry of $\sim$20 million stars with photometric
\feh\ and \afe\ calibrated on the spectra; two-epoch photometry of the Cepheids and Miras; and MIRI maps of the PAH and warm-dust
emission. We could apply the archeological methods and lessons learned from the Milky Way, on a second galaxy, distinct from our own.

\clearpage

\begin{multicols}{2}
\bibliography{M31_MCT_whitepaper_short}

@ARTICLE{2026arXiv260102520B,
       author = {{Barry}, Matthew and {Wetzel}, Andrew and {Loebman}, Sarah and others},
      journal = {arXiv:2601.02520},
         year = {2026}
}

@ARTICLE{2019A&A...631A..56B,
       author = {{Bhattacharya}, Souradeep and {Arnaboldi}, Magda and {Caldwell}, Nelson and others},
      journal = {\aap},
         year = {2019},
       volume = {631},
        pages = {A56}
}

@ARTICLE{2013ApJ...774...83B,
       author = {{Boyer}, Martha L. and {Girardi}, L{\'e}o and {Marigo}, Paola and others},
      journal = {\apj},
         year = {2013},
       volume = {774},
        pages = {83}
}

@ARTICLE{2019MNRAS.484.3476C,
       author = {{Clarke}, Adam J. and {Debattista}, Victor P. and {Nidever}, David L. and others},
      journal = {\mnras},
         year = {2019},
       volume = {484},
        pages = {3476-3490}
}

@ARTICLE{2023AJ....166...80D,
       author = {{Dalcanton}, Julianne J. and {Bell}, Eric F. and {Choi}, Yumi and others},
      journal = {\aj},
         year = {2023},
       volume = {166},
        pages = {80}
}

@ARTICLE{2015ApJ...803...24D,
       author = {{Dorman}, Claire E. and {Guhathakurta}, Puragra and {Seth}, Anil C. and others},
      journal = {\apj},
         year = {2015},
       volume = {803},
        pages = {24}
}

@ARTICLE{2018NatAs...2..737D,
       author = {{D'Souza}, Richard and {Bell}, Eric F.},
      journal = {Nature Astronomy},
         year = {2018},
       volume = {2},
        pages = {737-743}
}

@ARTICLE{2015ApJ...808..132H,
       author = {{Hayden}, Michael R. and {Bovy}, Jo and {Holtzman}, Jon A. and others},
      journal = {\apj},
         year = {2015},
       volume = {808},
        pages = {132}
}

@ARTICLE{2026arXiv261024915H,
       author = {{Hong}, Jihye and {Kirby}, Evan N. and {Gerasimov}, Roman and others},
      journal = {arXiv:2610.24915},
         year = {2026}
}

@ARTICLE{2020ApJ...889....5H,
       author = {{Huang}, Caroline D. and {Riess}, Adam G. and {Yuan}, Wenlong and others},
      journal = {\apj},
         year = {2020},
       volume = {889},
        pages = {5}
}

@ARTICLE{2023ApJ...956L..14K,
       author = {{Kobayashi}, Chiaki and {Bhattacharya}, Souradeep and {Arnaboldi}, Magda and others},
      journal = {\apjl},
         year = {2023},
       volume = {956},
        pages = {L14}
}

@ARTICLE{2024MNRAS.531..110K,
       author = {{Konchady}, Tarini and {Macri}, Lucas M. and {Yan}, Xiaomeng and others},
      journal = {\mnras},
         year = {2024},
       volume = {531},
        pages = {110-125}
}

@ARTICLE{2023ApJ...944L..17L,
       author = {{Lee}, Janice C. and {Sandstrom}, Karin M. and {Leroy}, Adam K. and others},
      journal = {\apjl},
         year = {2023},
       volume = {944},
        pages = {L17}
}

@ARTICLE{2015ApJ...805..183L,
       author = {{Lewis}, Alexia R. and {Dolphin}, Andrew E. and {Dalcanton}, Julianne J. and others},
      journal = {\apj},
         year = {2015},
       volume = {805},
        pages = {183}
}

@ARTICLE{2018MNRAS.477.5072M,
       author = {{Mackereth}, J. Ted and {Crain}, Robert A. and {Schiavon}, Ricardo P. and others},
      journal = {\mnras},
         year = {2018},
       volume = {477},
        pages = {5072-5089}
}

@ARTICLE{2019MNRAS.489..176M,
       author = {{Mackereth}, J. Ted and {Bovy}, Jo and {Leung}, Henry W. and others},
      journal = {\mnras},
         year = {2019},
       volume = {489},
        pages = {176-195}
}

@ARTICLE{2017AJ....154...94M,
       author = {{Majewski}, Steven R. and {Schiavon}, Ricardo P. and {Frinchaboy}, Peter M. and others},
      journal = {\aj},
         year = {2017},
       volume = {154},
        pages = {94}
}

@ARTICLE{1986A&A...154..279M,
       author = {{Matteucci}, F. and {Greggio}, L.},
      journal = {\aap},
         year = {1986},
       volume = {154},
        pages = {279-287}
}

@INPROCEEDINGS{2024IAUS..377..115N,
       author = {{Nidever}, David L. and {Gilbert}, Karoline and {Tollerud}, Erik and others},
    booktitle = {Early Disk-Galaxy Formation from JWST to the Milky Way},
       series = {IAU Symposium},
         year = {2024},
       volume = {377},
        pages = {115-122}
}

@ARTICLE{2018A&A...618A.156S,
       author = {{Saglia}, R.~P. and {Opitsch}, M. and {Fabricius}, M.~H. and others},
      journal = {\aap},
         year = {2018},
       volume = {618},
        pages = {A156}
}

@ARTICLE{2009MNRAS.396..203S,
       author = {{Sch{\"o}nrich}, Ralph and {Binney}, James},
      journal = {\mnras},
         year = {2009},
       volume = {396},
        pages = {203-222}
}

@ARTICLE{1979ApJ...229.1046T,
       author = {{Tinsley}, B.~M.},
      journal = {\apj},
         year = {1979},
       volume = {229},
        pages = {1046-1056}
}

@ARTICLE{2017ApJ...846..145W,
       author = {{Williams}, Benjamin F. and {Dolphin}, Andrew E. and {Dalcanton}, Julianne J. and others},
      journal = {\apj},
         year = {2017},
       volume = {846},
        pages = {145}
}

@ARTICLE{2023ApJS..268...48W,
       author = {{Williams}, Benjamin F. and {Durbin}, Meredith J. and {Dalcanton}, Julianne J. and others},
      journal = {\apjs},
         year = {2023},
       volume = {268},
        pages = {48}
}

@ARTICLE{2023ApJ...951...12W,
       author = {{Wojno}, Jennifer L. and {Gilbert}, Karoline M. and {Kirby}, Evan N. and others},
      journal = {\apj},
         year = {2023},
       volume = {951},
        pages = {12}
}

@ARTICLE{2007A&A...467..123B,
       author = {{Ballero}, S.~K. and {Matteucci}, F. and {Origlia}, L. and others},
      journal = {\aap},
         year = 2007,
       volume = {467},
        pages = {123},
       adsurl = {https://ui.adsabs.harvard.edu/abs/2007A&A...467..123B}
}

@ARTICLE{2018ARA&A..56..223B,
       author = {{Barbuy}, B. and {Chiappini}, C. and {Gerhard}, O.},
      journal = {\araa},
         year = 2018,
       volume = {56},
        pages = {223},
       adsurl = {https://ui.adsabs.harvard.edu/abs/2018ARA&A..56..223B}
}

@ARTICLE{2017A&A...605A..89B,
       author = {{Bensby}, T. and {Feltzing}, S. and {Gould}, A. and others},
      journal = {\aap},
         year = 2017,
       volume = {605},
        pages = {A89},
       adsurl = {https://ui.adsabs.harvard.edu/abs/2017A&A...605A..89B}
}

@ARTICLE{2018MNRAS.481.3210B,
       author = {{Bla{\~n}a D{\'\i}az}, M. and {Gerhard}, O. and {Wegg}, C. and others},
      journal = {\mnras},
         year = 2018,
       volume = {481},
        pages = {3210},
       adsurl = {https://ui.adsabs.harvard.edu/abs/2018MNRAS.481.3210B}
}

@ARTICLE{2016ARA&A..54..529B,
       author = {{Bland-Hawthorn}, J. and {Gerhard}, O.},
      journal = {\araa},
         year = 2016,
       volume = {54},
        pages = {529},
       adsurl = {https://ui.adsabs.harvard.edu/abs/2016ARA&A..54..529B}
}

@ARTICLE{2020MNRAS.491.5435B,
       author = {{Buck}, T.},
      journal = {\mnras},
         year = 2020,
       volume = {491},
        pages = {5435},
       adsurl = {https://ui.adsabs.harvard.edu/abs/2020MNRAS.491.5435B}
}

@ARTICLE{2008ApJ...684.1110C,
       author = {{Clarkson}, W. and {Sahu}, K. and {Anderson}, J. and others},
      journal = {\apj},
         year = 2008,
       volume = {684},
        pages = {1110},
       adsurl = {https://ui.adsabs.harvard.edu/abs/2008ApJ...684.1110C}
}

@ARTICLE{2021A&A...647A.131G,
       author = {{Gajda}, G. and {Gerhard}, O. and {Bla{\~n}a}, M. and others},
      journal = {\aap},
         year = 2021,
       volume = {647},
        pages = {A131},
       adsurl = {https://ui.adsabs.harvard.edu/abs/2021A&A...647A.131G}
}

@ARTICLE{2025arXiv251115415L,
       author = {{La Barbera}, F. and {Vazdekis}, A. and {Matteucci}, F. and others},
      journal = {arXiv e-prints},
         year = 2025,
        pages = {arXiv:2511.15415},
       adsurl = {https://ui.adsabs.harvard.edu/abs/2025arXiv251115415L}
}

@ARTICLE{2015MNRAS.447.1096M,
       author = {{Marcon-Uchida}, M.~M. and {Matteucci}, F. and {Lanfranchi}, G.~A. and others},
      journal = {\mnras},
         year = 2015,
       volume = {447},
        pages = {1096},
       adsurl = {https://ui.adsabs.harvard.edu/abs/2015MNRAS.447.1096M}
}

@ARTICLE{1990ApJ...365..539M,
       author = {{Matteucci}, F. and {Brocato}, E.},
      journal = {\apj},
         year = 1990,
       volume = {365},
        pages = {539},
       adsurl = {https://ui.adsabs.harvard.edu/abs/1990ApJ...365..539M}
}

@ARTICLE{1994ApJS...91..749M,
       author = {{McWilliam}, A. and {Rich}, R.~M.},
      journal = {\apjs},
         year = 1994,
       volume = {91},
        pages = {749},
       adsurl = {https://ui.adsabs.harvard.edu/abs/1994ApJS...91..749M}
}

@ARTICLE{2018ApJ...863...16R,
       author = {{Renzini}, A. and {Gennaro}, M. and {Zoccali}, M. and others},
      journal = {\apj},
         year = 2018,
       volume = {863},
        pages = {16},
       adsurl = {https://ui.adsabs.harvard.edu/abs/2018ApJ...863...16R}
}

@ARTICLE{2024MNRAS.530.2972S,
       author = {{Sanders}, J.~L. and {Kawata}, D. and {Matsunaga}, N. and others},
      journal = {\mnras},
         year = 2024,
       volume = {530},
        pages = {2972},
       adsurl = {https://ui.adsabs.harvard.edu/abs/2024MNRAS.530.2972S}
}

@ARTICLE{2005ApJ...631..280B,
       author = {R. Bender and J. Kormendy and G. Bower and others},
      journal = {ApJ},
       volume = {631},
        pages = {280-300},
         year = {2005}
}

@ARTICLE{2016ApJ...824...42C,
       author = {N. Caldwell and A. J. Romanowsky},
      journal = {ApJ},
       volume = {824},
        pages = {42},
         year = {2016}
}

@ARTICLE{2019ApJ...871L...8F,
       author = {G. Fragione and F. Antonini and O. Y. Gnedin},
      journal = {ApJL},
       volume = {871},
        pages = {L8},
         year = {2019}
}

@ARTICLE{2022A&A...657L..12G,
       author = {{GRAVITY Collaboration} and R. Abuter and A. Amorim and others},
      journal = {A\&A},
       volume = {657},
        pages = {L12},
         year = {2022}
}

@ARTICLE{2018MNRAS.475.2754H,
       author = {F. Hammer and Y. B. Yang and J. L. Wang and others},
      journal = {MNRAS},
       volume = {475},
        pages = {2754-2767},
         year = {2018}
}

@ARTICLE{1996AJ....112.1487H,
       author = {W. E. Harris},
      journal = {AJ},
       volume = {112},
        pages = {1487},
         year = {1996}
}

@ARTICLE{2015ApJ...806...96L,
       author = {T. C. Licquia and J. A. Newman},
      journal = {ApJ},
       volume = {806},
        pages = {96},
         year = {2015}
}

@INPROCEEDINGS{2015IAUS..311...82S,
       author = {J. Sick and S. Courteau and J. -C. Cuillandre and others},
    booktitle = {Galaxy Masses as Constraints of Formation Models},
       series = {IAU Symposium},
       volume = {311},
        pages = {82-85},
         year = {2015}
}

@ARTICLE{2015ApJ...799..144K,
       author = {{Kodric}, Mihael and {Riffeser}, Arno and {Seitz}, Stella and {Snigula}, Jan and {Hopp}, Ulrich and {Lee}, Chien-Hsiu and {Goessl}, Claus and {Koppenhoefer}, Johannes and {Bender}, Ralf and {Gieren}, Wolfgang},
        title = "{The M31 Near-infrared Period-Luminosity Relation and its Non-linearity for {\ensuremath{\delta}} Cep Variables with 0.5 <= log (P) <= 1.7}",
      journal = {\apj},
         year = 2015,
        month = feb,
       volume = {799},
       number = {2},
          eid = {144},
        pages = {144},
          doi = {10.1088/0004-637X/799/2/144},
archivePrefix = {arXiv},
       eprint = {1405.5218},
 primaryClass = {astro-ph.GA},
       adsurl = {https://ui.adsabs.harvard.edu/abs/2015ApJ...799..144K}
}

@ARTICLE{2018AJ....156..130K,
       author = {{Kodric}, Mihael and {Riffeser}, Arno and {Hopp}, Ulrich and {Goessl}, Claus and {Seitz}, Stella and {Bender}, Ralf and {Koppenhoefer}, Johannes and {Obermeier}, Christian and {Snigula}, Jan and {Lee}, Chien-Hsiu and {Burgett}, W.~S. and {Draper}, P.~W. and {Hodapp}, K.~W. and {Kaiser}, N. and {Kudritzki}, R.-P. and {Metcalfe}, N. and {Tonry}, J.~L. and {Wainscoat}, R.~J.},
        title = "{Cepheids in M31: The PAndromeda Cepheid Sample}",
      journal = {\aj},
         year = 2018,
        month = sep,
       volume = {156},
       number = {3},
          eid = {130},
        pages = {130},
          doi = {10.3847/1538-3881/aad40f},
archivePrefix = {arXiv},
       eprint = {1806.07895},
 primaryClass = {astro-ph.GA},
       adsurl = {https://ui.adsabs.harvard.edu/abs/2018AJ....156..130K}
}

@ARTICLE{2026arXiv260930551H,
       author = {{Huang}, Caroline D. and {Marengo}, Massimo and {Capodagli}, Antonio and {Riess}, Adam G. and {Breuval}, Louise and {Casertano}, Stefano and {Whitelock}, Patricia A. and {Anand}, Gagandeep S. and {Smith}, Randall K. and {Yuan}, Wenlong},
        title = "{Breaking the AGB Color Degeneracy with Variability and H$_2$O: A Consistent Picture of AGB Chemistry, Dust, and the J-Region Across Environments}",
      journal = {arXiv e-prints},
         year = 2026,
        month = sep,
          eid = {arXiv:2609.30551},
        pages = {arXiv:2609.30551},
          doi = {10.48550/arXiv.2609.30551},
archivePrefix = {arXiv},
       eprint = {2609.30551},
 primaryClass = {astro-ph.GA},
       adsurl = {https://ui.adsabs.harvard.edu/abs/2026arXiv260930551H}
}

@ARTICLE{2016MNRAS.459.1170S,
       author = {{Scowcroft}, Victoria and {Seibert}, Mark and {Freedman}, Wendy L. and {Beaton}, Rachael L. and {Madore}, Barry F. and {Monson}, Andrew J. and {Rich}, Jeffrey A. and {Rigby}, Jane R.},
        title = "{The Carnegie Chicago Hubble Program: the mid-infrared colours of Cepheids and the effect of metallicity on the CO band-head at 4.6 {\ensuremath{\mu}}m}",
      journal = {\mnras},
         year = 2016,
        month = jun,
       volume = {459},
       number = {2},
        pages = {1170-1178},
          doi = {10.1093/mnras/stw628},
archivePrefix = {arXiv},
       eprint = {1603.03776},
 primaryClass = {astro-ph.SR},
       adsurl = {https://ui.adsabs.harvard.edu/abs/2016MNRAS.459.1170S}
}

@ARTICLE{2018ApJ...857...16G,
       author = {{Goudfrooij}, Paul},
        title = "{Dissolved Massive Metal-rich Globular Clusters Can Cause the Range of UV Upturn Strengths Found among Early-type Galaxies}",
      journal = {\apj},
         year = 2018,
        month = apr,
       volume = {857},
       number = {1},
          eid = {16},
        pages = {16},
          doi = {10.3847/1538-4357/aab553},
archivePrefix = {arXiv},
       eprint = {1804.00668},
 primaryClass = {astro-ph.GA},
       adsurl = {https://ui.adsabs.harvard.edu/abs/2018ApJ...857...16G}
}

@ARTICLE{2019MNRAS.483.3420P,
       author = {{Parikh}, Taniya and {Thomas}, Daniel and {Maraston}, Claudia and {Westfall}, Kyle B. and {Lian}, Jianhui and {Fraser-McKelvie}, Amelia and {Andrews}, Brett H. and {Drory}, Niv and {Meneses-Goytia}, Sofia},
        title = "{SDSS-IV MaNGA: local and global chemical abundance patterns in early-type galaxies}",
      journal = {\mnras},
         year = 2019,
        month = mar,
       volume = {483},
       number = {3},
        pages = {3420-3436},
          doi = {10.1093/mnras/sty3339},
archivePrefix = {arXiv},
       eprint = {1812.02753},
 primaryClass = {astro-ph.GA},
       adsurl = {https://ui.adsabs.harvard.edu/abs/2019MNRAS.483.3420P}
}

@ARTICLE{2017ApJ...841...68V,
       author = {{van Dokkum}, Pieter and {Conroy}, Charlie and {Villaume}, Alexa and {Brodie}, Jean and {Romanowsky}, Aaron J.},
        title = "{The Stellar Initial Mass Function in Early-type Galaxies from Absorption Line Spectroscopy. III. Radial Gradients}",
      journal = {\apj},
         year = 2017,
        month = jun,
       volume = {841},
       number = {2},
          eid = {68},
        pages = {68},
          doi = {10.3847/1538-4357/aa7135},
archivePrefix = {arXiv},
       eprint = {1611.09859},
 primaryClass = {astro-ph.GA},
       adsurl = {https://ui.adsabs.harvard.edu/abs/2017ApJ...841...68V}
}
\end{multicols}
\end{document}